# Shifting Narratives: A Longitudinal Analysis of Media Trends and Public Attitudes on Homelessness

**Akshay Irudayaraj[1]\*, Nathan Ye[2]\*, and Yash Chainani[2]\***

[1]Brown University    [2]University of California, Berkeley

akshay_irudayaraj@brown.edu, {nathanye, yashchainani}@berkeley.edu

(\* indicates equal contribution)

## ABSTRACT

Within the field of media framing, homelessness has been a historically under-researched topic. Framing theory states that the media's method of presenting information plays a pivotal role in controlling public sentiment toward a topic. The sentiment held towards homeless individuals influences their ability to access jobs, housing, and resources as a result of discrimination. This study analyzes the topic and sentiment trends in related media articles to validate framing theory within the scope of homelessness. It correlates these shifts in media reporting with public sentiment. We examine state-level trends in California, Florida, Washington, Oregon, and New York from 2015 to 2023. We utilize the GDELT 2.0 Global Knowledge Graph (GKG) database to gather article data and use X to measure public sentiment towards homeless individuals. Additionally, to identify if there is a correlation between media reporting and public policy, we examine the media's impact on state-level legislation. Our research uses Granger-causality tests and vector autoregressive (VAR) models to establish a correlation between media framing and public sentiment. We also use latent Dirichlet allocation (LDA) and GPT-3.5 (LLM-as-annotator paradigm) for topic modeling and sentiment analysis. Our findings demonstrate a statistically significant correlation between media framing and public sentiment, especially in states with high homelessness rates. We found no significant correlation between media framing and legislation, suggesting a possible disconnect between public opinion and policy-making. These findings reveal the broader impact of the media's framing decisions and delineate its ability to affect society.

## I. INTRODUCTION

    In the United States, a large homeless population has been present for decades. In recent years, this number was around 580,000 individuals that lack a permanent residence (Sousa et al., 2022), and even with continued public and political attention on a national level, this number has not changed significantly. Some local and state leaders have taken actions regarding homelessness, some of which have been punitive while others seek to assist homeless individuals (Truong, 2012). In the general public, the majority of people are compassionate towards homeless individuals, though there is disagreement on how much government support should be provided to homeless individuals (Tsai et al., 2019). Public sentiment towards homelessness has been changing over the years as policies are passed, homeless numbers in individual communities change, and media coverage shifts (Christen & Huberty, 2007). Christen and Huberty also found that coverage in the media has been a major factor influencing public sentiment, specifically by framing issues in specific ways to induce certain views in the public.

    Framing theory aims to address the role of the media in shaping public opinion, with framing referring "to the process by which people develop a particular conceptualization of an issue or reorient their thinking about an issue" (Chong & Druckman, 2007, pp. 104–106). Framing can manifest in several regards: the use of specific words or metaphors to induce a certain sentiment, which topics are highlighted or discussed more than others, and the lens through which a topic is discussed (e.g. economic, political, etc.). Oftentimes, within the framing, small changes within the presentation of an issue can compound to create noticeable shifts in public opinion (Chong & Druckman, 2007). Topics are often framed by the media to show a specific point of view. For homelessness specifically, they can frame articles in a way that emphasizes certain characteristics about homeless individuals. The media's framing of homelessness can fuel misinformation or create widespread stereotypes that influence the way people perceive the homeless and as a result, how they act towards them. This can ultimately lead to societal waves that limit the ability for homeless individuals to get jobs, housing, and resources.

Thus, understanding the specific boundaries of the media's role is critical; however, the relationship between the media, public opinion, and policy efforts is complicated. Agenda-setting theory provides a framework for this role of the media in controlling the public narrative on issues (Entman, 1993). It can be broken down into three key steps: story selection, story framing, and how stories relate to each other (McCombs, 2006). First, the media dictates issue salience through their decisions of what news stories to cover; what the public deems as the important issues of the present is dependent on what news organizations decide to disseminate. Further, the media plays a key role in public perception of topics through how they decide to tell the story they choose. Finally, the relationships between news articles play a foundational role in spreading to the public a broader, grand narrative of what the world is. Which entities are discussed, how they are discussed, and the relationships between them power the integral role media has in controlling public knowledge (McCombs & Valuenza, 2014). While also relying on agenda-setting theory, this project primarily focuses on framing theory: understanding the influence media framing has on opinion and policy.

A previous examination of media framing on homelessness found that negative stereotypes such as mental illness and substance abuse were frequently discussed when writing about homelessness (Truong, 2012). The study also found that when the media addressed policies regarding homelessness, they highlighted individual behaviors, such as panhandling, rather than structural inequities, such as an increase in the lack of affordable housing. The negative framing of the homeless led to punitive policies that aimed to criminalize homelessness, with policymakers framing the motivation of the policies around crime prevention. This lines up with framing theory, as the constant discussion of the homeless in a negative light shifts the public perception of the homeless negatively, and policymakers follow up by creating policies that punish the homeless rather than assist them. For example, from 2006 to 2019, city-wide bans on camping, loitering, and other activities associated with homelessness have significantly increased (National Homelessness Law Center, 2019). However, these laws fail to address the underlying causes of homelessness as homeless encampments grew by 1300% across the United States (Tars, 2021).

This study will analyze articles about homelessness from March 2015 to December 2023 to extract the main themes discussed in each article (e.g. drug use, veterans, subsidized housing), and the sentiment within the article towards homeless individuals. It will then determine the change in sentiment and frequency of specific topics in articles about homelessness over time. It will aim to determine the factors for which the change most correlates to a change in public sentiment towards homeless individuals. The objective of this is to identify how different topics discussed in the media can influence views towards homeless individuals, which can have an impact on the ability of homeless individuals to get jobs, access resources, and rebuild their lives. Additionally, this study will identify turning points in these topic and sentiment trends and attempt to correlate these turning points with new public policy regarding homelessness. Essentially, this research aims to answer:

1. Does the framing of homelessness in articles correlate to shifts in public opinion towards homeless individuals?
2. Does the shift in article framing (if any) correlate with state-level policy reforms?

The current study differentiates from any previous research done on the media framing of homelessness because it examines a far more comprehensive online news collection of articles relevant to homelessness, and analyzes state-wide impacts on public opinion and legislation. Legislation holds importance in that it defines the conditions for the homeless, and determining whether or not supportive legislation is passed is a key research insight. Additionally, previous research on media framing of the homeless such as those done within Canada (Calder et al., 2011) or San Francisco (Fries, 2020) have not analyzed digital news sources, instead primarily studying newspapers, radio, or television. When capturing the current media framing of homelessness, digital sources are a significant source of information as the majority of Americans receive their news from digital devices and news websites, which is what this study seeks to examine (Liedke & Wang, 2023).

It is important to understand the relationship between new organizations' framing of homelessness and shifts within public sentiment and legislation for several reasons. First, by extrapolating patterns from which framing contexts are the most influential, readers can begin to realize the role polarization plays in informing their opinions. Showcasing that especially sensationalized concepts or visceral emotions create the largest shifts in opinion would contribute to readers' knowledge of how article emotionality impacts their views. With this knowledge, they will be equipped to refocus their immediate reactions, perhaps with a more logical perspective. Second, an examination of what context homeless people are framed in contributes to the public's understanding of which negative stereotypes are impactful, both from a personal perspective and policy one. Uncovering what leads the public to embrace punitive policies rather than supportive ones is especially relevant considering the controversy over the City of Grants Pass vs. Johnson SCOTUS case that furthers the criminalization of the homeless. Finally, learning more about the strength of the correlation between public sentiment and legislation provides key insights into the effectiveness of the US democracy. Adding on the work of Page (1994) and others, we can analyze whether the large influence of media priming and independent government action follows in the realm of homelessness.

## II. METHODOLOGY

**A. Datasets**

One of the major parts of our analysis was to identify the trends in topics discussed in articles related to homelessness and correlate the discussion of topics in the media with shifts in public sentiment. In order to analyze the trends in topics and sentiment related to articles about homelessness, we used the Global Database of Events, Language, and Tone (GDELT) Global Knowledge Graph (GKG) database. GKG is a dataset that allows us to view the topics, emotions, and locations of nearly all news articles published by recognized news sources around the world. To filter by articles related to homelessness, we only used articles that contain the GDELT-labeled topics "need of shelters'' and "poverty." These filters yielded a total of 825,000 articles. As our analysis was on a state-by-state basis, we used the locations provided by GDELT to filter out articles that didn't include the state of interest. Rather than a nationwide analysis, we used state-level analyses since sentiments towards homelessness can be very different based on political views, incidence, and discussion of homelessness within each state, and thus should be analyzed at a state level. We used the states California, New York, Florida, Washington, and Oregon as these were among the US states with the highest homeless populations.

Another part of our analysis involved identifying the trend over time in public sentiment toward homeless individuals. To get the sentiment information, we scraped X (formerly known as Twitter) using Selenium. X queries contain multiple ways to specify the type of posts returned. For each of the states we studied, we limited the examined X posts using the "near: [state]" advanced search technique. Additionally, we iterated over each week from January 1st, 2015 to December 31st, 2023 using the "before: [date]" and "until: [date]" search techniques. For each state and week, we utilized a boolean, phrase searched query: ("homeless" or "homelessness" or "unhoused"). For each of these queries, Selenium allowed us to gather the text data for each post as well as the exact date of each post. Note that this scraping method was not uniform, in that there was some variability (though not very significant) across the number of tweets scraped for each query due to X's harsh restrictions on bot scraping. In total, the number of tweets we gathered amounted to around 14,000 per state (or 70,000 total). While scraping was less robust, it allowed us to expand the scope and computational extensiveness of our project due to X's recently implemented limits to its application programming interface (API).

Finally, to capture the impact on public policy, our study required the collection of state legislation that involves homelessness in California, Oregon, Florida, Washington, and New York. Specifically, the text of any legislation discussed from 2015 to 2023 was scraped off official state government websites, which includes any state bills and amendments brought up in state government sessions. To be comprehensive, we examined both the state senate and state house legislation, searching for the keywords "homeless", "unhoused", or "unsheltered", which are keywords used to describe the homeless. From these keywords, we gathered 4,523 pieces of legislation in total. For the timeframe of legislation, we expanded the scope to a yearly basis, as it was often unclear when legislation was first brought up and to account for varying times for legislation to be introduced or passed. Thus the year of each piece of legislation was also included within our data collection process. However, the states of California, New York, and Washington recorded their legislation within the 2-year terms for state legislators, so the specific year was not displayed on the website when scraping. Given several hundred pieces of legislation were discussed every year, it was impractical to manually label the years of each piece of legislation. Instead, the large language model GPT-3.5 Turbo was used to annotate the year of discussion (see TABLE 2 column 1 in the Appendix section for prompt). GPT's accuracy was verified with a manual verification for years on a simple random sample of 55 pieces of legislation relevant to homelessness, ending up with an acceptable accuracy of 0.87.

**B. Variables**

From the filtered articles by state from GDELT, we identified the number of articles published per month for each topic. Furthermore, we divided the count for each topic into two sections: articles with a positive sentiment and articles with a negative sentiment based on whether the Linguistic Inquiry and Word Count (LIWC) positive emotion or negative emotion dictionary evaluation was greater (Boyd et. al, 2022). Adding these factors allowed our analysis to better consider the influence of each topic by grouping articles with a similar narrative towards a topic together, rather than mixing together opposite sentiments. Finally, these topic-sentiment combination counts were normalized by dividing the raw number for each topic in a month by the total number of articles in the month. This accounts for variation in the total number of articles per month and allows us to view the trend in the topics chosen by the media regarding homelessness rather than being influenced by trends in the total number of articles, and total number of sources writing about homelessness. This process allowed us to identify trends over the years of how frequently each homelessness-related topic was discussed in the media, which we then correlated with public sentiment towards homeless individuals.

Once we gathered our X posts, we chose to classify them based on sentiment using the LLM-as-annotator paradigm with GPT-3.5 Turbo. Using OpenAI's Batch API, we fed these posts to GPT-3.5 with the query listed in TABLE 2. Critically, we used few-shot prompting here, so in each prompt, we listed an example for each of the scores we asked it to output. These options were: 1: strongly anti-homeless, 2: slightly anti-homeless, 3: neutral to homeless/unrelated to homeless individuals, 4: slightly pro-homeless, 5: strongly pro-homeless. Note also that we instructed the system with some context that it would be given an X post and that it should rate the post's sentiment based purely on its attitude towards homeless individuals, not the overall sentiment.

To analyze legislation trends, we needed to identify the main legislative topics discussed within each state for every year. Since the time frame was set on a yearly basis, all the legislation text for each state was fed into a latent Dirichlet allocation (LDA) model to identify the top 9 topics, and keywords related to each of these topics (Blei et. al, 2003). We decided on 9 since it optimized the topic coherence value ($C_v$) for a sample of legislation years (Mimno et. al, 2011). Since LDA does not specify the exact topic being covered, it required further annotation to transform LDA-generated keywords into GKG themes before analysis between state legislative topics and media GKG themes could be performed. Due to the large number of LDA topics and there being over 800,000 GKG themes, the most feasible form of annotation was through using the LLM-as-annotator paradigm. Thus, we prompted GPT-3.5 Turbo (see TABLE 2, column 2 of the Appendix section for prompt) to annotate LDA topics into the top three GKG themes for each topic. We fed GPT-3.5 the top 500 GKG themes that were discussed in the media articles we analyzed related to homelessness, as it was unfeasible to pass in over 800 thousand themes. To test intercoder reliability feasibly, as long as our manual coders annotated a topic under one of the GKG themes, we considered it a match. The Krippendorff's alpha between manual coders and GPT was calculated to be 0.643.

**C. Analysis**

To analyze the relationship between individual topic-sentiment frequencies and public opinion, we used Granger-causality tests, since the test was able to determine if the topic-sentiment time series could forecast the X sentiment time series (Granger, 1969). There were a series of filtration steps we had to run in order to ensure the data was processed correctly and also as necessary prerequisites to run the Granger-causality test. First, we removed all infrequently mentioned topic-sentiment frequencies (any that were mentioned on average in below 1.5% of all articles). We decided on the 1.5% cut-off value experimentally (looking at values slightly above and below) since it ensured that barely mentioned topics (that were mentioned for example, many times in just one month) did not skew the data while also retaining the primary topics salient to discussions of homelessness. Additionally, a necessary assumption of the Granger-causality test is stationarity, so we had to ensure that the topic-sentiment time series we passed into the test exhibited consistent statistical properties. To do this, we performed an augmented Dickey-Fuller (ADF) test, which extends on the general Dickey-Fuller test for stationarity to account for higher order autocorrelations and lagged differences (Dickey & Fuller, 1979). With these prerequisites met, we were able to perform the Granger-causality tests between individual topic-sentiment time series and public opinion. Notably, we solely focused on the sum of squares (SSR) based F-statistic provided from the test and used that to calculate the *p*-value since the degrees of freedom (df) for our test were quite low (at max 3) and our null hypothesis was that there was no relationship. We iterated over a lag from 1 month to 3 months, stopping there since we believed that any correlations past that point would be due to random chance alone. Also, along with running a Granger-causality test between the topic-sentiment time series and public opinion time series, we calculated a cross-correlation coefficient, which provides us with the linear correlation between the two variables accounting for the given lag. To uncover the relationship between multiple topic-sentiment time series and public opinion, we had to look beyond Granger-causality and decided on a vector autoregressive (VAR) model since it was well-equipped to handle our multivariate use case (Sims, 1980). Before we passed the most significant topics into the VAR model, we first removed multicollinear columns via a principal component analysis (PCA), which allowed to condense redundant columns into one (Pearson, 1901). Note that a prerequisite to applying PCA was a standard scaling process to normalize the data. We chose the optimal number of components for the PCA such that the sum of the variances of the components added up to a standard value of 95%; in other words, we ensured that the chosen number of components explained 95% of the dataset's variance.

Next, since retaining the features' original names was critical, we used PCA loadings to reconstruct which of the original component names most contributed to the filtration products. In order to meet the prerequisites to run the VAR test, we could only pass in 6 features in order to maximize the observation to feature ratio across a max lag of 1. We chose the 6 most explanatory and correlated variables from our filtered data and passed these into the VAR model. From this, we were able to determine if multiple topic-sentiment time series had a statistically significant ability to forecast public opinion. In order to determine the correlation between these values, we decided on a ordinary least squares (OLS) regression due to its easy interpretability and robustness as opposed to alternatives like averaging out the cross-correlation coefficients of the individual topics (Gauss, 1821).

In order to highlight the connection between media framing and legislation, we analyzed if topics being discussed in the context of homelessness more within media had a relationship with the same topics appearing within legislation for each state. For each state, we separated the yearly relative frequencies for every GKG theme into two different groups: years where the GKG theme appeared within state legislation and years where it did not appear in its respective state's legislation. Then, we ran two-sample $t$-tests for every GKG theme within every state to determine if there were statistically significant differences between how much a GKG theme is mentioned within GDELT articles in years where they were discussed in legislation and years where they were not discussed. It is important to note that the GKG theme had to appear in at least two years within each state and also not be mentioned within two years of legislation, as otherwise, the sample size was too small for a $t$-test. Additionally, to analyze salience in the greater context of homelessness, we also ran linear regression tests between the total GDELT article count for each state as well as the legislation count for each state from 2015 to 2023, examining a yearly basis.

## III. RESULTS

### A. Media Framing and X Sentiment

From each state, through the rigorous selection process described in the methodology, the twenty most prominent GDELT topic-sentiment combinations were identified and listed in TABLES 3-7 (in the Appendix), with each table showcasing these important topics for each state. Here, statistical significance ($p < 0.05$) indicates that the shifts in individual topic-sentiment discussion forecast (or Granger-cause) shift in X sentiment; in other words, the analyzed topic follows a proportional relationship (positive or negative) offset by the minimal lag. Considering the topic-sentiment combination as a predictor of sentiment shifts as opposed to considering them separately via the VAR model, a statistically significant was established since the $p$-values for all the states were under 0.05. However, the $r$ and $r^2$ values, represented in TABLE 1, exhibited a somewhat weak correlation and a low amount of sentiment variance explained, respectively.

**TABLE 1.** VAR and Regression Outputs per State

| state | $p$ | $r$ | $r^2$ |
|---|---|---|---|
| CA | 0.002 | 0.402 | 0.162 |
| OR | 0.024 | 0.179 | 0.032 |
| WA | < 0.001 | 0.395 | 0.156 |
| FL | 0.001 | 0.346 | 0.112 |
| NY | < 0.001 | 0.512 | 0.262 |

**FIG. 1.** The Top 6 GDELT GKG Themes in New York Averaged and Compared with Average Sentiment Against Date

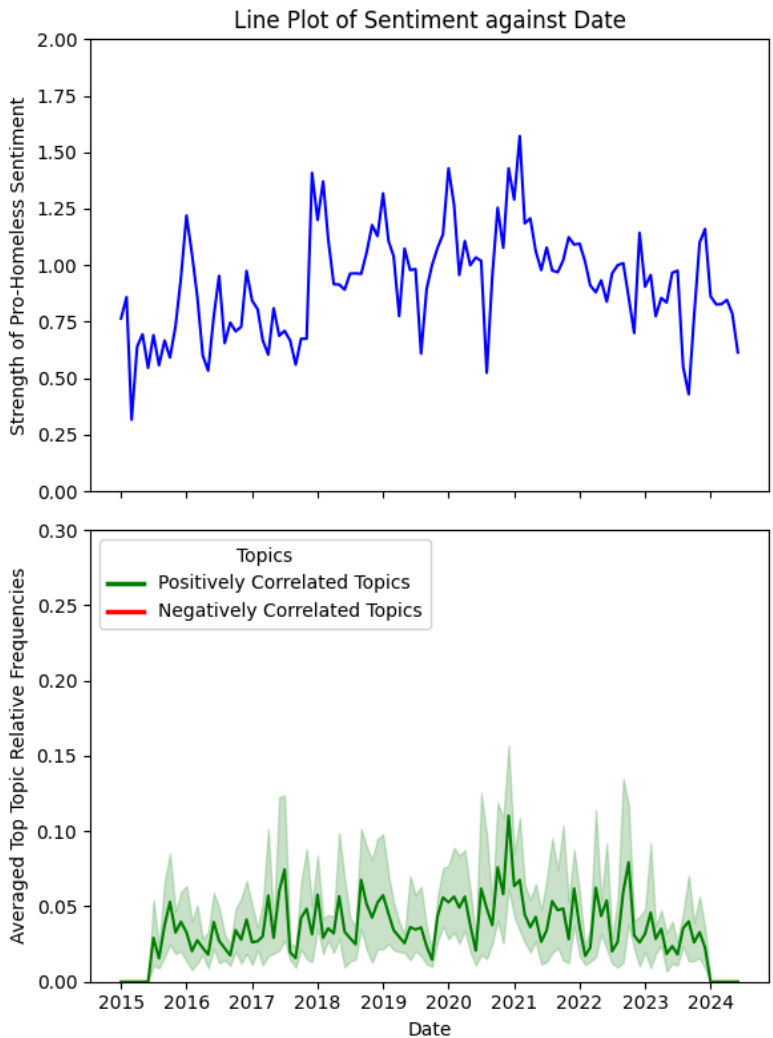

In FIG. 1, the relative percentages of the top 6 GDELT GKG themes in New York are compared against sentiment longitudinally. The two follow a generally similar trend where increased discussions of the most positively correlated topics correspond (though offset by the average lag) with spikes in X sentiment, supporting the conclusions gathered from the Granger causality test. These trends within each state's graph depend on their respective correlation values, and they can be viewed in FIG. 4-7 (in the Appendix).

**B. Media Framing and Legislation**

Based on TABLES 3-7 in the Appendix section, none of the two-sample t-tests for all states and GKG themes ran between the relative frequency of GKG themes in articles when they were mentioned in legislation and not mentioned in legislation revealed a statistically significant relationship ($p < 0.05$). These t-tests show no relationship between GKG themes appearing in articles and in legislation. Additionally in FIG. 2, correlating article count per year and bill count per year reveals no statistically significant relationship ($p = 0.604$) and an extremely small correlation. Overall, no statistically significant relationships within the analysis done for this research question were found.

**Fig. 2** Article Count vs. Bill Count Per Year

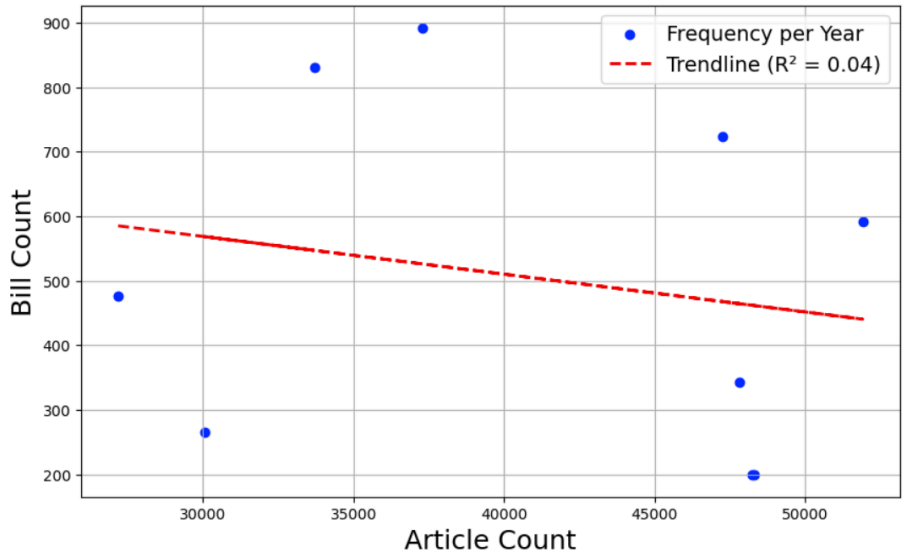

*Note*. This figure examines all articles and bills analyzed based on the total counts analyzed. Similar insignificant relationships were found within all states examined.

### C. Homelessness Salience in Media

**FIG. 3.** Homelessness Article Count Per Unique Source

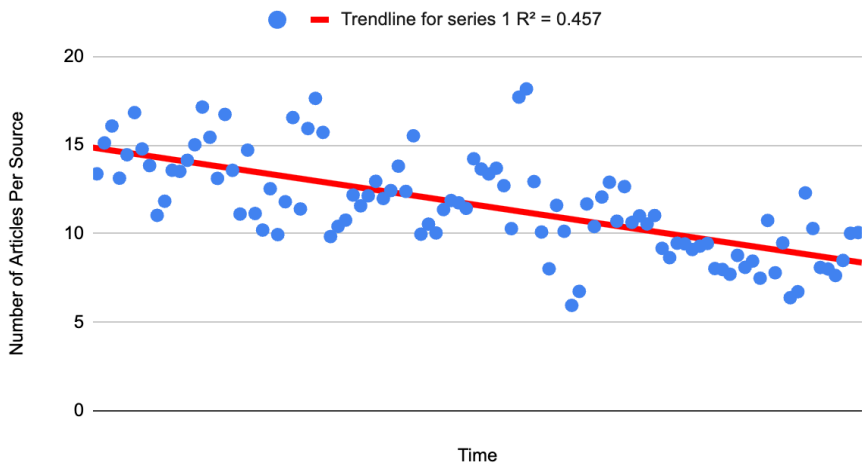

  An unexpected finding of the study was that the overall discussion of homelessness in the media has declined over time, as shown in FIG. 3. Since the number of sources varied over different months, we looked at the average number of articles per source, per month rather than the raw article count. From this, we found that the number of articles per source related to homelessness has been declining over time, effectively demonstrating that the media is caring less and less about addressing issues related to homelessness.

## IV. DISCUSSION

  Based on our state-level analysis, many themes relevant to homelessness have a statistically significant effect on the average monthly X sentiment, which suggests a relationship between these themes and public attitudes toward the homeless. The overarching categories of topics we identified across states as being correlated with public sentiment were government, social, economic, health, and education. Government, social, economic, and health topics were positively correlated with public sentiment, while education was negatively correlated with sentiment. Within these categories, some of the sub-topics were discussed in a positive sentiment while others were discussed negatively.

Within the category of government, the main subtopics were congressional policy and local legislation (both positive and negative article sentiments had a positive correlation with public sentiment), as well as taxes (negative article sentiment had a positive correlation with public sentiment). This correlation may be due to people viewing legislation and tax policies as not doing enough for the homeless. As a result, they feel more sympathy for these individuals and thus have a more positive sentiment.

For the social category, some of the main topics we identified as having a positive correlation with sentiment were veterans, disabled persons, and human rights. This makes sense based on societal views towards these groups. Veterans are generally viewed positively by the public, and people have sympathy for them. Associating homelessness with veterans can increase sentiment as people realize the homeless individuals they see could be veterans suffering from the after-effects of a war such as PTSD. Persons with disabilities also have greater sympathy from the general population. Disabilities can often be viewed as out of a person's control. This can increase sentiment towards homeless individuals, in contrast to stereotypes that suggest that it is within a person's control to exit homelessness.

Positive correlations were also observed with topics related to economics, particularly those highlighting unemployment and a struggling economy. The increased coverage of these economic hardships may lead to greater public sympathy toward the homeless, as it frames their struggles within the context of broader economic challenges. By emphasizing systemic issues such as job scarcity and economic instability, the media can influence public perception, leading people to view homelessness as a consequence of structural factors rather than individual failings.

Health-related topics such as access to medical services, and the health of disabled individuals, also correlated with a positive sentiment. Once again, this reflects a factor out of the control of individuals and rather, a consequence of societal structure. People feel sympathy towards homeless individuals without access to healthcare and can have a more positive sentiment as a result.

A commonality between the topics that lead to positive sentiments is that the majority of them can be seen as out of the control of homeless individuals. In contrast, the topics that lead to negative sentiments can be viewed as within the control of these individuals. For instance, education leads to a negative correlation since it may be seen as within a person's control to obtain an education up to high school with free public education, and possibly lead to the public blaming the homeless for their own circumstances. Other topics we found to lead to negative sentiments included drugs and crime, which may be due to the association of the homeless with these negative topics and also perpetuate negative stereotypes of them. Overall, these findings imply that individual topics themselves have an impact on the public perception of homelessness, and they can lead to a positive or negative sentiment toward the homeless due to the role they play in framing homelessness. Please note that these interpretations are a qualitative analysis we underwent based on the quantitative results we gathered.

As discussed in the Results section, we established a weak but statistically significant relationship for Granger-causality between the most influential topic-sentiment combinations and shifts in public sentiment scores for each state, which reinforces framing theory within the realm of homelessness. The framing of news articles related to homelessness does affect public opinion, though at a small level. However, within social science, correlation coefficients above 0.1 can be understood as acceptance since there are so many factors behind the dependent variables (Ozili, 2022). Within our research, the weak correlation can be explained by the variety of other factors affecting public opinion, such as social group biases and traditions. Notably, the tone in which topics were discussed was significant in addition to topic salience. We noticed that the positive and negative variants of the topics did not both show up as significant — from this, we can infer that for media sources, it's not enough that they discuss certain topics but the way they frame them has differing impacts. Additionally, we uncovered differences in the correlations between states. States with the highest homelessness rates, like New York, exhibited noticeably higher correlation values. A potential reason for this may be that in states where homelessness is more present to the general public, there is likely increased media reporting on the issue, and people might be more inclined to pay attention to it since it manifests in their daily lives.

In our second research question, only salience was examined, and whether or not homelessness or any of their subtopics were discussed more, they had no effect on the type or amount of state legislation discussed. Based on our first findings, the topics discussed by the media have an effect on the people but do not affect legislation, which suggests that legislators may not have the same priorities as the public. This brings negative implications on democracy on a state level, as rather than caring about what the public thinks on the topic of homelessness, legislators may be discussing bills based on their personal agendas or other factors rather than the people putting them in power.

This study has several key limitations that may impact the scope of its findings. First, the source of public sentiment is not representative of all Americans in a state who consume news media, as it relies on X users who post on homelessness. Additionally, only X's geolocation filter did not completely limit posts to a specified state and allowed posts from nearby states to be included. Furthermore, the locations of GDELT articles were also not limited to any one specific state, as many articles had multiple different location tags. The articles within the GDELT database can also be accessed by media viewers in any state online, which means that other states' media could have affected media viewers in any specific state being examined. Thus, our operationalizations for media and public sentiment were not representative of their states.

Based on examining LDA-generated words, a significant amount of legislation may not have been directly relevant and skewed the LDA topical analysis towards irrelevant topics, which made it more difficult to track homelessness subtopics focused in the media and subtopics addressed by legislation. Additionally, LDA performs better on text data with fewer words, so topic modelling methods that take into account context (such as BERTopic) may have been a more effective topic modelling method.

This study also relied on large language models, particularly OpenAI's ChatGPT 3.5, to annotate large datasets, as previous studies have attempted as well (Chae & Davidson, 2023). For coding X sentiments toward homelessness, a high intercoder reliability of Krippendorff's alpha of 0.802 was achieved. However, the LLM was also prompted to generate GKG themes based on LDA topic words, many of which did not describe subtopics of homelessness, and were far less accurate with an intercoder reliability of 0.643, which reduces the reliability of findings relevant to subtopics of homelessness addressed in legislation. A reason for this may be the complexity of the prompts and type of prompting, since the first prompt required little token parsing and was few-shot while the second required the model to read through the large list of GKG themes and was zero-shot. Generally, the LLM-as-annotator paradigm was effective in our study only for relatively simple tasks and with guidance.

Finally, this study's Granger causality tests had relatively large lags of one to three months as our analysis was limited to monthly analyses due to limited X scraping capabilities. These large time durations could potentially have limited the correlations of our analysis, as X users may react within a month and sentiments may change much faster than the time gaps we analyzed. Additionally, articles may also have different levels of influence on public perception based on how many views they receive, engagement, and other factors that were not weighed within this study.

Future research can be done on homelessness with a more representative source of public sentiment, which can be done through a repeated survey that measures public opinion towards the homeless and the potential existence of stereotypes within the public. Using a representative survey will allow findings to be more generalized and have a larger scope, especially if more states beyond the five in the current study are examined. As for legislation, rather than using LDA and LLM coding, a more reliable topic classification approach could be through crowdsourced manual coding, which would identify topics more relevant to homelessness for a more accurate analysis. To avoid missing out on changes within small timeframes, future analyses should look at a weekly or even daily timeframe to potentially find greater relationships. Furthermore, measures not used within this study such as morality or other measures of sentiment can be done on the GDELT database to see if other factors may have an impact on public sentiment towards homelessness, especially given the low correlations we found between any individual variable and public sentiment.

The paper contributes to the confirmation of framing theory in the under researched topic of homelessness, as we demonstrate how the media frames the issue, specifically in salience and valence, impacts public opinion. Additionally, it demonstrates the viability of the LLM-as-annotator paradigm, especially for simple tasks such as labeling years or sentiments towards the homeless when given examples with the few-shot method, as we found high intercoder reliability for those tasks. As for more complex tasks with much more subjectivity, such as defining topics from LDA-generated words, LLM is far less accurate, so we recommend that future researchers use a codebook and more examples to ensure higher accuracy and intercoder reliability. Our findings also demonstrate to the general public that they should be critical of media portrayals of homelessness and other social issues, understanding that the framing used by the media can significantly influence their perceptions and attitudes. Overall, seeking multiple perspectives and being aware of potential biases in media reporting is instrumental in finding an accurate portrayal of the homeless.

## ACKNOWLEDGEMENTS

We would like to thank our SRA Track 3 instructor, Musa Malik, as well as our course TAs, Brittany Wheeler and Sovannie Len. Additionally, thank you to Dr. Lina Kim and Dr. Teresa Holden for organizing the SRA program and providing us with the opportunity to conduct this research.

## AUTHOR CONTRIBUTION STATEMENT

## Appendix

Disclosure Statement: We acknowledge the use of GPT-3.5 Turbo (https://platform.openai.com/docs/models/gpt-3-5-turbo) to annotate state legislation years, sentiment toward the homeless within X posts, and transform LDA keywords into GKG themes. The prompts used can be seen in TABLE 2.

The output from these prompts is explained in TABLE 2.

**TABLE 2.** ChatGPT Prompts along with their output use

| ChatGPT Prompts | Reason |
|---|---|
| return the year (either session year 1 or year 2) the bill was introduced based on this excerpt of the bill, ONLY the number and no text, if it cannot be located, return the closest year with ONLY the number and it can ONLY be [session year 1] or [session year 2]" | GPT was necessary to annotate the year of each piece of legislation, as there were far too many to manually annotate and they were not identified within the website. |
| You will be given a list of 10 related words. These are words that an LDA model extracted from state bills regarding homelessness. Please pick subtopics from the list of subtopics provided below. You can choose up to 3 subtopics. Do not provide any additional description: just have 3 subtopics from the list below separated by commas. HARD DISCLAIMER: PLEASE TRY TO ANSWER WITH A SUBTOPIC OF HOMELESSNESS, rather than just broadly describing homelessness. Here is an example categorization: (list of words: custody inmate adult department person correction service court board subsection program public employee rule read, Output topics: SOC_GENERALCRIME, ARREST, SOC_POINTSOFINTEREST_PRISON) Here is the list of subtopics from which you must choose (some of them have irrelevant prefixes; you can ignore the prefix in these cases): [list of all GKG themes] | GPT was necessary to transform LDA topic summaries into GKG themes. |
| You will be given a post on Twitter. Analyze the post and give me a category for its attitude towards homeless individuals. You are specifically analyzing sentiment towards the homeless, not the general sentiment of the post. (1. Strongly anti-homeless, 2. Somewhat anti-homeless, 3. Neutral to homeless/unrelated to homeless individuals, 4. Somewhat pro-homeless, 5. Strongly pro-homeless). Just return a number and nothing else (1, 2, 3, 4, or 5). Here is an example of each category: Category 1: Can U put out a video letting us know what 2 do 2 protect ourselves from the insane/drugged out homeless men on the streets? 2. Category 2: Your city or state might be one Democratic vote away from your own homeless encampment on wheels! Category 3: Cities across California and the nation are experiencing a rise in homelessness Category 4: They'd rather cover the city in spikes like it's bowser's | GPT was the optimal way to label the sentiment towards homeless individuals in Twitter posts for us to analyze sentiment trends (directed sentiment towards homelessness rather than just overall). |

castle than build homeless shelters Category 5: I REALLY hate the way people treat the homeless. Can you imagine how it feels to be treated with such disdain day in & day out?

**FIG. 4.** The Top 6 GDELT GKG Themes in California Averaged and Compared with Average Sentiment Against Date

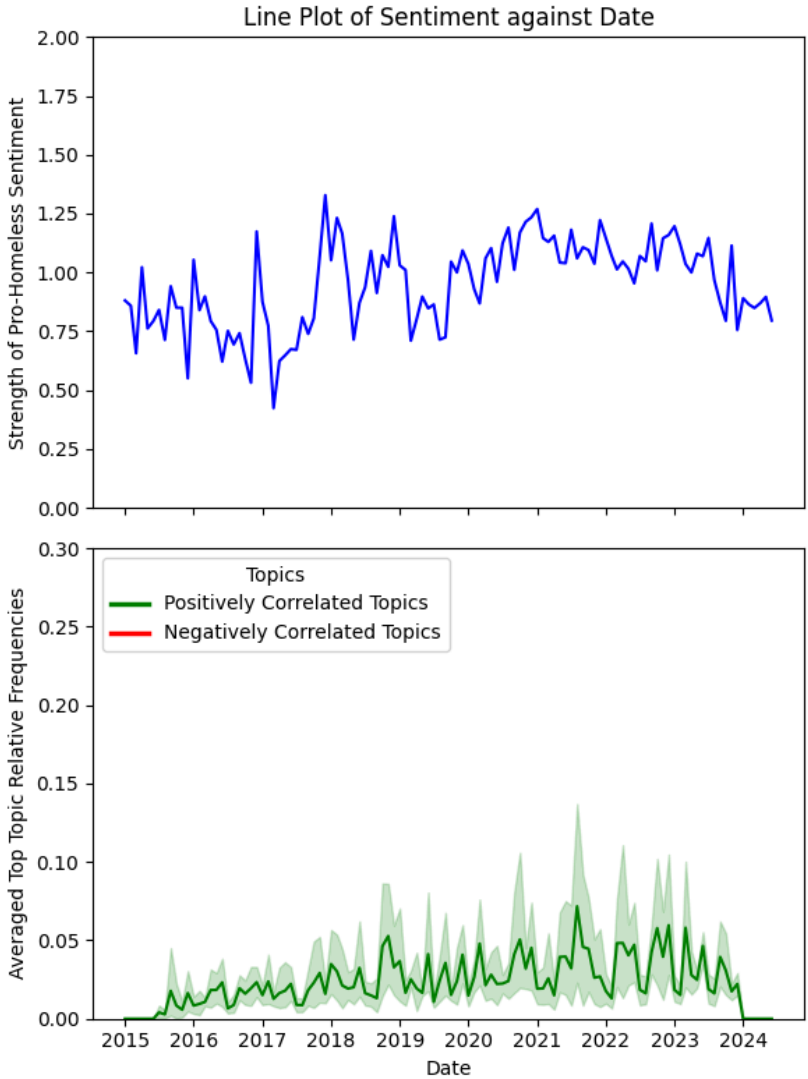

**FIG. 5.** The Top 6 GDELT GKG Themes in Oregon Averaged and Compared with Average Sentiment Against Date

12 of 21

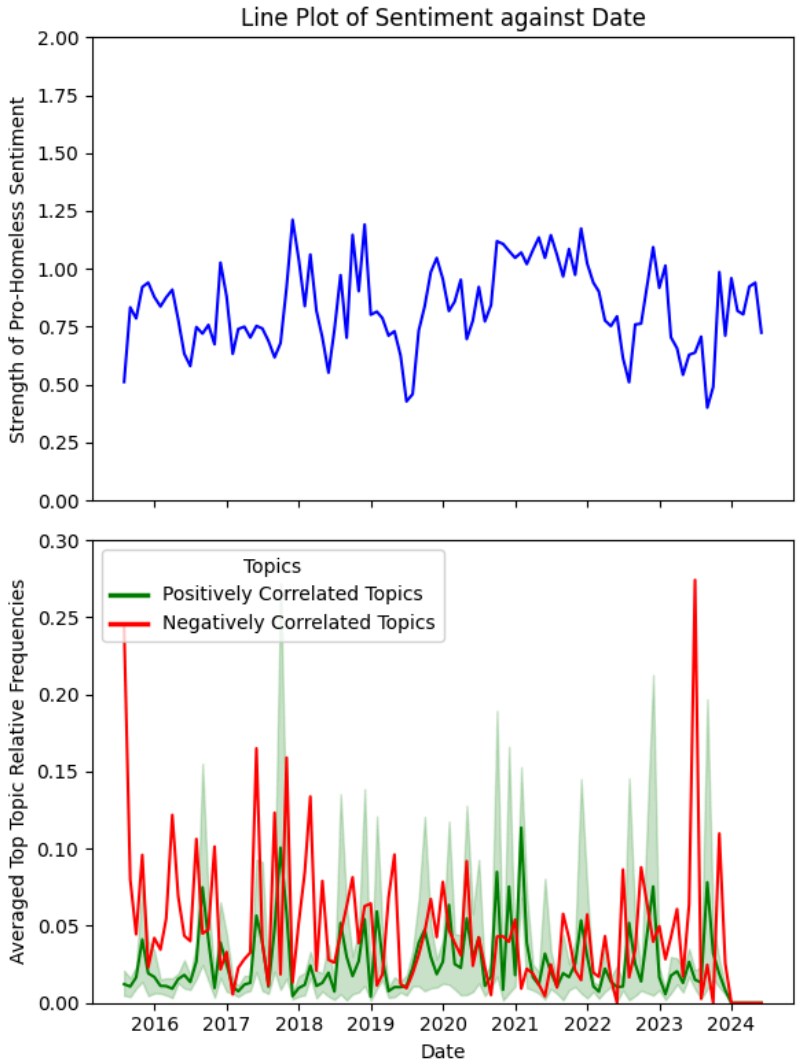

**FIG. 6.** The Top 6 GDELT GKG Themes in Washington Averaged and Compared with Average Sentiment Against Date

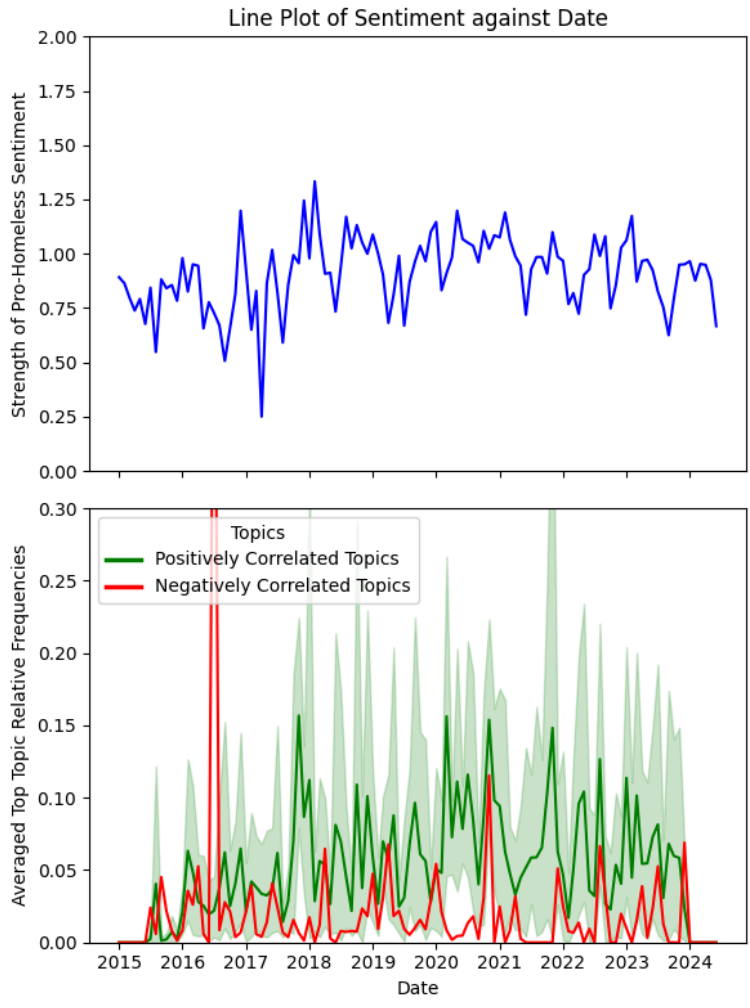

**FIG. 7.** The Top 6 GDELT GKG Themes in Florida Averaged and Compared with Average Sentiment Against Date

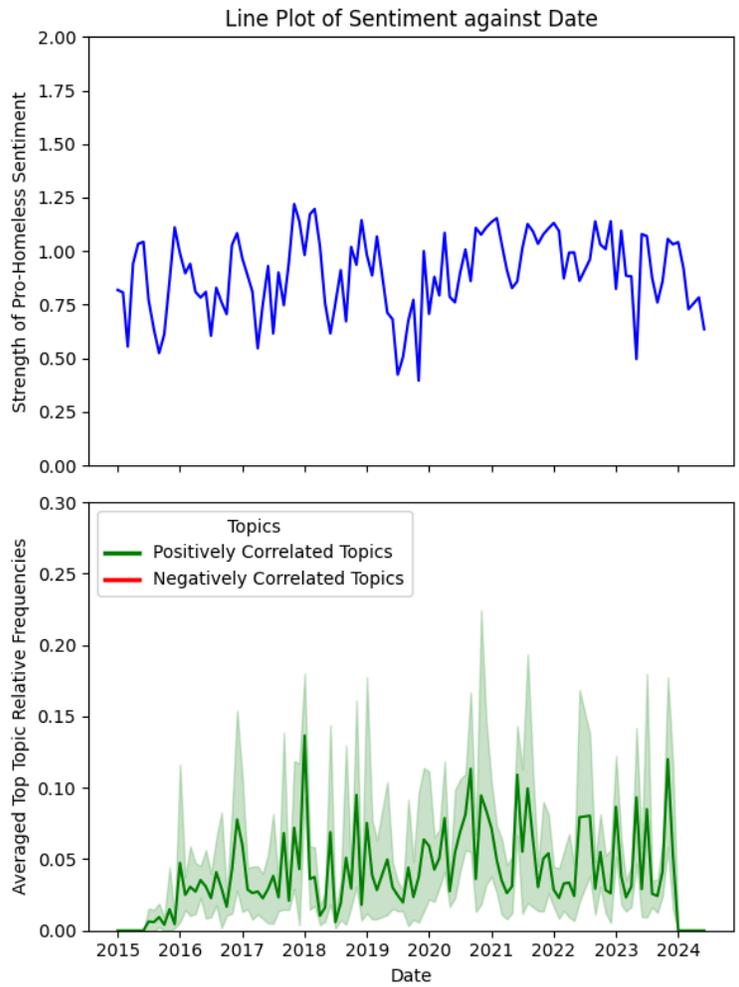

**TABLE 3.** Twenty Most Correlated Topics that Were Significant Within California

| Topic | cc | p | Lag |
|---|---|---|---|
| EPU_POLICY_CONGRESSIONAL_POS | 0.358 | 0.007 | 1 |
| EPU_CATS_TAXES_NEG | 0.332 | 0.030 | 1 |
| EPU_UNCERTAINTY_POS | 0.298 | 0.041 | 1 |
| WB_1464_HEALTH_OF_THE_DISABLED_POS | 0.287 | 0.006 | 1 |
| ECON_STOCKMARKET_POS | 0.278 | 0.018 | 1 |
| DISPLACED_NEG | 0.263 | 0.038 | 2 |
| TAX_FNCACT_ASSISTANT_POS | 0.260 | 0.019 | 1 |
| TAX_FNCACT_SECRETARY_NEG | 0.257 | 0.037 | 1 |
| WB_1305_HEALTH_SERVICES_DELIVERY_NEG | 0.239 | 0.050 | 1 |
| WOUND_POS | 0.229 | 0.016 | 1 |
| WATER_SECURITY_NEG | -0.224 | 0.030 | 3 |
| TAX_FNCACT_SHERIFF_POS | -0.219 | 0.047 | 2 |
| TAX_FNCACT_WORKER_POS | 0.216 | 0.040 | 1 |
| TAX_FNCACT_CHILDREN_NEG | 0.213 | 0.034 | 1 |
| EPU_CATS_MIGRATION_FEAR_MIGRATION_NEG | 0.204 | 0.045 | 2 |
| STRIKE_POS | 0.188 | 0.047 | 1 |
| CRISISLEX_T04_INFRASTRUCTURE_POS | 0.155 | 0.046 | 2 |
| WB_168_ROADS_AND_HIGHWAYS_POS | -0.138 | 0.048 | 3 |
| TAX_WORLDMAMMALS_FOX_POS | -0.089 | 0.039 | 1 |
| TAX_FNCACT_FATHER_POS | -0.046 | 0.003 | 2 |

*Note*. cc refers to the correlation coefficient accounting for lag and the p-value refers to the result of the individual Granger-causality test. The following are true for TABLES 4-7.

**TABLE 4.** Twenty Most Correlated Topics that Were Significant Within Oregon

| Topic | cc | p | Lag |
|---|---|---|---|
| DISPLACED_POS | 0.272 | 0.034 | 3 |
| TAX_FNCACT_STUDENT_POS | -0.255 | 0.044 | 3 |
| TAX_FNCACT_MANAGER_NEG | 0.250 | 0.002 | 1 |
| ECON_HOUSING_PRICES_NEG | 0.248 | 0.049 | 2 |
| SOC_USSECURITYAGENCIES_POS | 0.232 | 0.002 | 2 |
| SELF_IDENTIFIED_HUMAN_RIGHTS_POS | 0.198 | 0.047 | 1 |
| TAX_FNCACT_DEPUTY_NEG | 0.192 | 0.040 | 1 |
| POLITICAL_TURMOIL_NEG | 0.190 | 0.046 | 2 |
| WB_1428_INJURY_NEG | -0.184 | 0.035 | 1 |
| DISPLACED_NEG | 0.181 | 0.043 | 1 |
| TAX_FNCACT_ATTORNEY_POS | -0.163 | 0.016 | 2 |
| DRUG_TRADE_NEG | -0.155 | 0.037 | 1 |
| EPU_POLICY_GOVERNMENT_NEG | 0.142 | 0.029 | 1 |
| TAX_FNCACT_DRIVERS_POS | -0.140 | 0.033 | 1 |
| WB_698_TRADE_POS | 0.124 | 0.042 | 2 |
| WB_1199_WATER_SUPPLY_AND_SANITATION_POS | -0.118 | 0.049 | 2 |
| TAX_ECON_PRICE_POS | -0.104 | 0.008 | 2 |
| CRISISLEX_O01_WEATHER_POS | -0.101 | 0.034 | 1 |
| TAX_FNCACT_POLICE_OFFICER_POS | -0.087 | 0.018 | 2 |
| SOC_POINTSOFINTEREST_HIGH_SCHOOL_POS | -0.053 | 0.042 | 3 |

**TABLE 5.** Twenty Most Correlated Topics that Were Significant Within Washington

| Topic | cc | p | Lag |
|---|---|---|---|
| UNGP_JOB_OPPORTUNITIES_EMPLOYMENT_NEG | 0.334 | 0.018 | 3 |
| UNGP_CRIME_VIOLENCE_POS | 0.312 | 0.008 | 1 |
| TAX_FNCACT_KING_NEG | 0.306 | 0.001 | 1 |
| TAX_DISEASE_OUTBREAK_NEG | 0.303 | 0.010 | 1 |
| WB_478_LEVELS_OF_EDUCATION_POS | -0.289 | 0.011 | 2 |
| EPU_POLICY_CONGRESS_NEG | 0.287 | 0.045 | 1 |
| TRANSPARENCY_NEG | -0.284 | 0.007 | 1 |
| SHORTAGE_NEG | 0.274 | 0.015 | 2 |
| WB_2745_JOB_QUALITY_AND_LABOR_MARKET_PERFORMANCE_NEG | 0.255 | 0.015 | 1 |
| GOV_LOCALGOV_POS | 0.246 | 0.004 | 2 |
| ENV_OIL_POS | 0.227 | 0.041 | 2 |
| MILITARY_NEG | 0.223 | 0.025 | 1 |
| TAX_FNCACT_CONGRESSMAN_POS | 0.211 | 0.043 | 1 |
| USPEC_UNCERTAINTY1_NEG | 0.210 | 0.044 | 1 |
| WB_1462_WATER_SANITATION_AND_HYGIENE_POS | 0.205 | 0.029 | 1 |
| PROTEST_POS | 0.199 | 0.049 | 2 |
| ARMEDCONFLICT_POS | 0.196 | 0.026 | 1 |
| UNGP_EDUCATION_POS | -0.188 | 0.020 | 2 |
| TAX_FNCACT_LAWYER_POS | 0.179 | 0.026 | 3 |
| UNGP_FORESTS_RIVERS_OCEANS_NEG | 0.136 | 0.016 | 1 |

**TABLE 6.** Twenty Most Correlated Topics that Were Significant Within Florida

| Topic | cc | p | Lag |
|---|---|---|---|
| CORRUPTION_POS | 0.335 | 0.019 | 2 |
| EPU_POLICY_CONGRESS_POS | 0.309 | 0.009 | 1 |
| UNGP_JOB_OPPORTUNITIES_EMPLOYMENT_NEG | 0.268 | 0.015 | 1 |
| EPU_CATS_ENTITLEMENT_PROGRAMS_NEG | 0.267 | 0.001 | 1 |
| USPEC_UNCERTAINTY1_NEG | 0.256 | 0.024 | 1 |
| STATE_OF_EMERGENCY_NEG | 0.238 | 0.011 | 1 |
| TAX_FNCACT_VETERANS_POS | 0.233 | 0.027 | 1 |
| WATER_SECURITY_NEG | 0.226 | 0.008 | 2 |
| LGBT_NEG | -0.213 | 0.046 | 3 |
| EPU_POLICY_CONGRESS_NEG | 0.212 | 0.004 | 1 |
| WB_1199_WATER_SUPPLY_AND_SANITATION_NEG | 0.209 | 0.047 | 2 |
| SOC_POINTSOFINTEREST_HOSPITAL_POS | 0.209 | 0.046 | 1 |
| CRIME_ILLEGAL_DRUGS_POS | -0.207 | 0.012 | 1 |
| TAX_FNCACT_SHERIFF_NEG | -0.196 | 0.029 | 1 |
| SOC_POINTSOFINTEREST_SCHOOLS_POS | 0.183 | 0.045 | 1 |
| TAX_FNCACT_FATHER_NEG | 0.183 | 0.015 | 1 |
| TAX_DISEASE_DEPRESSION_NEG | 0.150 | 0.043 | 1 |
| WB_167_PORTS_NEG | 0.144 | 0.021 | 2 |
| EPU_CATS_TAXES_NEG | -0.122 | 0.024 | 3 |
| TAX_FNCACT_DOCTOR_POS | 0.030 | 0.019 | 3 |

**TABLE 7.** Twenty Most Correlated Topics that Were Significant Within New York

| Topic | cc | p | Lag |
|---|---|---|---|
| TAX_FNCACT_DOCTOR_POS | 0.417 | 0.000 | 1 |
| SOC_POINTSOFINTEREST_HEADQUARTERS_POS | 0.414 | 0.000 | 2 |
| APPOINTMENT_POS | 0.371 | 0.000 | 1 |
| TAX_FNCACT_CEO_POS | 0.341 | 0.010 | 2 |
| TAX_FNCACT_COMMISSIONER_POS | 0.340 | 0.039 | 2 |
| TAX_WORLDLANGUAGES_MASSACHUSETTS_POS | 0.280 | 0.049 | 3 |
| TRANSPARENCY_POS | 0.274 | 0.033 | 3 |
| TAX_WORLDLANGUAGES_RUSSIA_POS | 0.260 | 0.028 | 2 |
| TAX_FNCACT_VETERANS_POS | 0.248 | 0.034 | 1 |
| WB_507_ENERGY_AND_EXTRACTIVES_NEG | 0.239 | 0.006 | 3 |
| RAPE_NEG | 0.224 | 0.010 | 1 |
| NATURAL_DISASTER_HURRICANE_POS | 0.217 | 0.000 | 1 |
| TAX_FNCACT_ATTORNEYS_NEG | 0.195 | 0.023 | 1 |
| REBELLION_POS | 0.186 | 0.034 | 3 |
| TAX_FNCACT_REPRESENTATIVE_POS | 0.185 | 0.012 | 3 |
| STATE_OF_EMERGENCY_NEG | 0.176 | 0.015 | 1 |
| ELECTION_NEG | 0.171 | 0.010 | 1 |
| TAX_FNCACT_WRITER_NEG | 0.139 | 0.002 | 2 |
| TAX_FNCACT_OFFICIAL_NEG | 0.127 | 0.041 | 1 |
| DRUG_TRADE_POS | -0.061 | 0.047 | 1 |

**TABLE 8.** *t*-Tests Between Relative Frequency of Articles with GKG Themes When Mentioned In Legislation and Not Mentioned in Legislation for California

| GKG Theme | t-statistic | p-value | Mentioned | | Not Mentioned | |
|---|---|---|---|---|---|---|
| | | | mean | sd | mean | sd |
| GENERAL_HEALTH | -.71 | .52 | .54 | .16 | .62 | .17 |
| MEDICAL | 1.11 | .35 | .54 | .06 | .48 | .10 |
| CRISISLEX_CRISISLEXREC | -.27 | .80 | .78 | .04 | .78 | .03 |
| SECURITY_SERVICES | .30 | .77 | .37 | .01 | .36 | .05 |
| WB_635_PUBLIC_HEALTH | 1.94 | .10 | .41 | .23 | .22 | .04 |
| WB_621_HEALTH_NUTRITION_AND_POPULATION | .61 | .64 | .64 | .22 | .55 | .13 |
| LEGISLATION | 1.83 | .16 | .34 | .03 | .29 | .03 |
| GOV_LOCALGOV | .72 | .49 | .12 | .01 | .11 | .03 |
| SOC_POINTSOFINTEREST_SCHOOL | -.83 | .43 | .22 | .03 | .24 | .03 |
| HEALTH_VACCINATION | 1.06 | .36 | .09 | .12 | .03 | .03 |
| EPU_POLICY_AUTHORITIES | 1.12 | .38 | .20 | .04 | .16 | .05 |
| EPU_POLICY_BUDGET | -.12 | .91 | .09 | .01 | .09 | .05 |

**TABLE 9.** *t*-Tests Between Relative Frequency of Articles with GKG Themes When Mentioned In Legislation and Not Mentioned in Legislation for Florida

| GKG Theme | t-statistic | p-value | Mentioned | | Not Mentioned | |
|---|---|---|---|---|---|---|
| | | | mean | sd | mean | sd |

| GKG Theme | t-statistic | p-value | Mentioned mean | Mentioned sd | Not Mentioned mean | Not Mentioned sd |
|---|---|---|---|---|---|---|
| SOC_POINTSOFINTEREST | .03 | .97 | .69 | .02 | .69 | .03 |
| CRISISLEX_CRISISLEXREC | .00 | 1.00 | .80 | .02 | .80 | .06 |
| WB_697_SOCIAL_PROTECTION_AND_LABOR | .55 | .61 | .21 | .02 | .20 | .06 |
| GENERAL_GOVERNMENT | .35 | .74 | .38 | .10 | .36 | .08 |
| WB_470_EDUCATION | -1.09 | .39 | .28 | .05 | .33 | .06 |
| URBAN | -2.09 | .08 | .17 | .04 | .23 | .05 |
| WB_2432_FRAGILITY_CONFLICT_AND_VIOLENCE | 1.45 | .34 | .46 | .05 | .40 | .03 |
| WB_635_PUBLIC_HEALTH | -1.01 | .38 | .23 | .11 | .35 | .21 |
| TAX_FNCACT_CHILDREN | -1.36 | .37 | .29 | .04 | .33 | .02 |
| TAX_DISEASE | -.19 | .88 | .41 | .36 | .46 | .13 |
| MEDICAL | .81 | .45 | .52 | .10 | .48 | .06 |
| MEDIA_MSM | -.62 | .56 | .30 | .00 | .31 | .04 |
| WB_2495_DETENTION_PRISON_AND_CORRECTIONS_REFORM | 2.35 | .11 | .16 | .03 | .11 | .02 |
| CRIME_ILLEGAL_DRUGS | 2.25 | .06 | .08 | .01 | .06 | .02 |
| ARREST | -.57 | .59 | .28 | .02 | .29 | .05 |
| SOC_POINTSOFINTEREST_SCHOOL | .47 | .68 | .31 | .10 | .29 | .02 |
| WB_699_URBAN_DEVELOPMENT | -.29 | .78 | .07 | .02 | .07 | .03 |
| TAX_FNCACT_CHILD | .61 | .56 | .15 | .02 | .14 | .03 |
| DISCRIMINATION | -.64 | .55 | .08 | .03 | .09 | .02 |
| ECON_DEVELOPMENTORGS | -1.71 | .13 | .01 | .01 | .02 | .02 |
| CRIME_COMMON_ROBBERY | 1.99 | .15 | .05 | .01 | .04 | .01 |
| WB_1305_HEALTH_SERVICES_DELIVERY | .17 | .89 | .05 | .04 | .05 | .01 |
| HEALTH_VACCINATION | 1.73 | .22 | .15 | .13 | .02 | .01 |
| HEALTH_PANDEMIC | 1.22 | .38 | .35 | .24 | .12 | .22 |
| EPU_POLICY_GOVERNMENT | -1.71 | .21 | .18 | .14 | .33 | .07 |
| UNGP_EDUCATION | 1.14 | .32 | .03 | .01 | .02 | .01 |

**TABLE 10.** *t*-Tests Between Relative Frequency of Articles with GKG Themes When Mentioned In Legislation and Not Mentioned in Legislation for Oregon

| | | | Mentioned | | Not Mentioned | |
|---|---|---|---|---|---|---|
| GKG Theme | t-statistic | p-value | mean | sd | mean | sd |
| SOC_POINTSOFINTEREST | 1.37 | .22 | .63 | .05 | .59 | .04 |
| LEGISLATION | .35 | .74 | .35 | .08 | .32 | .10 |
| CRISISLEX_CRISISLEXREC | -1.87 | .14 | .76 | .06 | .82 | .03 |
| GENERAL_HEALTH | .85 | .43 | .63 | .13 | .54 | .19 |
| SOC_POINTSOFINTEREST_SCHOOL | .77 | .48 | .26 | .04 | .24 | .03 |
| CRIME_ILLEGAL_DRUGS | -2.00 | .12 | .07 | .02 | .11 | .04 |
| MEDICAL | .00 | 1.00 | .51 | .17 | .51 | .08 |
| ARREST | .40 | .70 | .29 | .02 | .28 | .08 |
| TAX_FNCACT_CHILDREN | -.38 | .73 | .20 | .06 | .22 | .04 |

| GKG Theme | | | | | | |
|---|---|---|---|---|---|---|
| SECURITY_SERVICES | -2.27 | .10 | .35 | .03 | .41 | .05 |
| TAX_FNCACT_POLICE | .07 | .95 | .35 | .02 | .35 | .05 |
| BAN | -1.73 | .25 | .10 | .05 | .17 | .05 |
| EPU_POLICY_POLITICAL | -.21 | .84 | .11 | .02 | .11 | .07 |
| HEALTH_PANDEMIC | .28 | .82 | .22 | .31 | .16 | .22 |

**TABLE 11.** *t*-Tests Between Relative Frequency of Articles with GKG Themes When Mentioned In Legislation and Not Mentioned in Legislation for New York

| | | | Mentioned | | Not Mentioned | |
|---|---|---|---|---|---|---|
| **GKG Theme** | **t-statistic** | **p-value** | mean | sd | mean | std |
| TAX_FNCACT_MAN | -.34 | .77 | .31 | .06 | .33 | .06 |
| MANMADE_DISASTER_IMPLIED | -.65 | .54 | .73 | .01 | .75 | .04 |
| TAX_FNCACT_POLICE | -.32 | .77 | .33 | .05 | .35 | .03 |
| CRIME_ILLEGAL_DRUGS | 2.82 | .13 | .11 | .02 | .08 | .02 |
| WB_695_POVERTY | 1.70 | .25 | .23 | .03 | .18 | .04 |
| LEGISLATION | .82 | .52 | .35 | .03 | .33 | .02 |
| WB_1809_HIGHWAYS | 1.42 | .20 | .05 | .00 | .05 | .02 |
| GENERAL_HEALTH | -1.52 | .18 | .51 | .08 | .65 | .18 |
| SOC_POINTSOFINTEREST_SCHOOL | .93 | .39 | .30 | .01 | .28 | .04 |
| TAX_FNCACT_OFFICIALS | -.54 | .67 | .34 | .08 | .37 | .05 |
| ARREST | 1.51 | .22 | .31 | .03 | .27 | .03 |
| GENERAL_GOVERNMENT | .33 | .75 | .37 | .05 | .35 | .06 |
| WB_470_EDUCATION | -1.80 | .19 | .25 | .04 | .32 | .06 |
| WB_840_JUSTICE | -1.86 | .17 | .31 | .04 | .35 | .02 |
| TAX_FNCACT_CHILDREN | -.71 | .50 | .30 | .02 | .31 | .04 |
| TAX_FNCACT_POLITICIANS | .80 | .55 | .07 | .02 | .06 | .01 |
| WB_635_PUBLIC_HEALTH | -.30 | .79 | .31 | .18 | .35 | .21 |
| TAX_FNCACT_CHILD | -.87 | .44 | .13 | .01 | .14 | .02 |
| ECON_DEVELOPMENTORGS | .62 | .56 | .03 | .01 | .02 | .02 |
| HEALTH_PANDEMIC | .89 | .43 | .29 | .27 | .13 | .23 |
| EPU_POLICY_GOVERNMENT | -1.20 | .30 | .21 | .14 | .30 | .06 |

**TABLE 12.** *t*-Tests Between Relative Frequency of Articles with GKG Themes When Mentioned In Legislation and Not Mentioned in Legislation for Washington

| | | | Mentioned | | Not Mentioned | |
|---|---|---|---|---|---|---|
| **GKG Theme** | **t-statistic** | **p-value** | mean | sd | mean | sd |
| GENERAL_HEALTH | -1.42 | .20 | .59 | .12 | .71 | .13 |
| SOC_POINTSOFINTEREST | -.68 | .52 | .68 | .04 | .71 | .07 |
| LEGISLATION | .69 | .51 | .38 | .06 | .36 | .01 |
| SOC_POINTSOFINTEREST_SCHOOL | -.44 | .69 | .31 | .07 | .33 | .04 |
| TAX_FNCACT_CHILDREN | .87 | .41 | .30 | .00 | .29 | .03 |
| SECURITY_SERVICES | -.53 | .67 | .34 | .08 | .37 | .05 |
| TAX_FNCACT_LANDLORD | 1.02 | .37 | .04 | .01 | .03 | .01 |
| GOV_LOCALGOV | -1.97 | .09 | .11 | .01 | .14 | .04 |
| HEALTH_PANDEMIC | .01 | .99 | .18 | .31 | .18 | .22 |